\documentclass[aps,prb,showpacs,showkeys]{revtex4}

\usepackage[T2A,OT1]{fontenc}
\usepackage[cp1251]{inputenc}         
\usepackage[russian,ukrainian,english]{babel} 
\usepackage[dvips]{epsfig}            
\usepackage{latexsym}
\usepackage{amsfonts}
\usepackage{boxedminipage}

\renewcommand{\theenumi}{\arabic{enumi}}

\begin{document}
\selectlanguage{english} 
%
\title{Magnetoelastic nature of the solid oxygen $\varepsilon$-phase structure}
\author{H. V. Gomonay  and V. M. Loktev}
\date{\today}
\affiliation
{Bogolyubov Institute for Theoretical Physics NAS of
Ukraine,\\ Metrologichna str. 14-b, 03143, Kyiv, Ukraine}
\begin{abstract}  For a long time a crystal
structure of the high-pressure $\varepsilon$-phase of solid oxygen
was a mystery. Basing on the results of recent experiments that
have solved this riddle it is shown that the magnetic and crystal
structure of this phase can be explained by strong exchange
interactions of antiferromagnetic nature. The singlet state
implemented on quaters of O$_2$ molecules has the minimal exchange
energy if compared to other possible singlet states (dimers,
trimers). Magnetoelastic forces that arise from the space
dependence of the exchange integral give rise to transformation of
4(O$_2$) rhombuses into the almost regular quadrates.
Antiferromagnetic character of exchange interactions stabilizes
the distortion of crystal lattice in $\varepsilon$-phase and
impedes such a distortion in the long-range $\alpha$- and
$\delta$-phases.
\end{abstract}

 \maketitle

 Solid oxygen is known to occupy a particular place
in the family of cryocrystals \cite{Cryocrystals_e}. Steady
interest to oxygen during almost 50 years is due to the magnetic
properties of O$_2$ molecule which posses nonzero spin $S_{O_2}=1$
in the ground electronic state. This ensures magnetism of all the
solid O$_2$ phases.

Solid oxygen has a rather complicated phase diagram, which
includes three low temperature phases stable at ambient pressure:
$\gamma$ ($T<43.8$~K), $\beta$ ($23.8\le T\le 43.8$~K) and
$\alpha$ ($T\le 23.8$~K), and several high pressure phases:
$\delta$, $\varepsilon$ and $\zeta$ (see review
\cite{Freiman:2004} which includes also a comprehensive history of
a problem). All these phases except $\zeta$-O$_2$ are insulators.
Metallization of solid O$_2$ takes place at 96 GPa
\cite{Desgreniers:1990, Akahama:1995} at room temperature.
Moreover, $\zeta$-O$_2$ was also observed in a superconducting
state \cite{Shimizu:1998}.

An interesting feature of all (except
 $\gamma$-O$_2$) the phases of solid oxygen is
parallel alignment of the molecules which is usually explained by
strong contribution of exchange interactions into anisotropic
(i.e. depending on the mutual orientation of molecules) part of
intermolecular potential. Presence of stable orientation ordering
enables to simplify substantially many of theoretical models and,
in particular, makes it possible to describe the structural phase
transitions in solid oxygen disregarding orientational dynamics of
molecules.

In the absence of the orientational degrees of freedom, the
low-temperature rhombohedral (space group $R\overline{3}m$)
$\beta$-phase can be thought of as a para-phase for all the
magnetic phases. In particular, $\beta$-O$_2$ has a planar
structure, consisting of close packing of parallel oxygen
molecules with centers of mass in the apexes of regular triangles,
and with orientation perpendicular to the basal planes. The
temperature dependence of magnetic susceptibility of $\beta$-O$_2$
is typical for antiferromagents (AFM)
\cite{Borovik-Romanov:1954E}. Noncollinear 3-sublattice ordering
in this phase (Loktev structure) was predicted in
\cite{Loktev:1979E} and now is generally accepted. Below 23.9 K
(or at high pressure) $\beta$-phase becomes unstable and
transforms into monoclinic (space group C2/m) $\alpha$-phase.
Corresponding $\alpha\beta$-phase transition has magnetoelastic
nature associated with strong dependence of exchange interaction
vs intermolecular distance, as it was shown in
\cite{Loktev:1981E}. The $\alpha$-phase possesses the collinear
(N\'eel) magnetic structure with the easy direction parallel to
the monoclinic axis $\mathbf{b}$ of slightly distorted (compared
to regular hexagonal) lattice. Long-range AFM ordering, which can
be described within a simple 2-sublattice model, is stabilized by
the ``deformation-induced splitting'' of intra- and
inter-sublattice exchange integrals.  It should be stressed that
all the in-plane exchange constants originate from a single
constant $J(\mathbf{r})$ that has AFM character (i.e.,
$J(\mathbf{r})>0$), is isotropic and describes intermolecular spin
interactions in $\beta$-phase \cite{Loktev:1975E}. Mutual shift of
the close-packed basal planes that accompanies formation of AFM
ordering also has magnetoelastic nature and originates from space
dependence of inter-plane exchange integral \cite{gomo:2005}.

Hydrostatic pressure up to 4$\div$6~GPa induces continuous shift
of the basal planes, while the magnetic structure of
$\alpha$-O$_2$ and orientation of molecules remain invariable. At
approximately 6.5~GPa the mutual shift of neighboring planes
attains 1/2 of an in-plane intermolecular distance and solid O$_2$
transforms into orthorhombic (space group Fmmm) $\delta$-phase.
The type of magnetic order in $\alpha$- and $\delta$-phases is
similar (collinear AFM structure) within $ab$-plane, but relative
orientation of spins in the neighboring planes (between the first
interplane neighbors) is different -- parallel in $\alpha$-O$_2$
and antiparallel in $\delta$-O$_2$ \cite{Goncharenko:2004}. Due to
crucial change of magnetic structure (from 2 to 4-sublattice)
$\alpha\delta$-phase transition is classified as the Ist
order\footnote{~In the absence of magnetic interactions
$\alpha\delta$-transition can be classified as a IInd order,
according to Liftshits' criterium.}. Abrupt change of magnetic
order at the $\alpha\delta$-transition point is accompanied by
discontinuous shift of the close-packed planes, which also
originates from space dependence of the inter-plane exchange
integral.

Further increase of pressure up to 8~GPa produces another phase
transformation
 into $\varepsilon$-phase which elusive structure
has been determined recently \cite{Lundegaard:2006,
Fujihisa:2006}. The transition is undoubtedly of the Ist order and
is accompanied by a considerable (up to 5.4\%) volume reduction.
Crystal structure of $\varepsilon$-phase is layered, as it is the
case for $\alpha$-, $\beta$-, and $\delta$-phases, and has
monoclinic (space group C2/m) symmetry. Variation of interplane
distance (equal $\approx 3.4$~\AA at the transition point) with
pressure is very small. So, the volume change is mainly due to the
variation of intermolecular distances within the basal plane. The
peculiar feature of $\varepsilon$-phase is association of four
O$_2$ molecules into rhomb-shaped (according to
Ref.\onlinecite{Lundegaard:2006}) or square-shaped  (according to
Ref.\onlinecite{Fujihisa:2006}) (O$_2$)$_4$ molecular units, which
are symmetry equivalent and centered on the lattice points at
(0,0,0) and (0.5,0.5,0). Common spin-state of the (O$_2)_4\equiv$
O$_8$ cluster is nonmagnetic\footnote{~Disappearence of the
magnetic properties of solid oxygen was experimentally proved in
Ref.\onlinecite{Goncharenko:2005}.}, with the total spin
$S_{O_8}=0$.

Physical reasons of such an unusual behavior of the magnetic
molecular crystal are not yet clearly understood. Should
$\varepsilon$-phase be considered as a chemically new substance,
what is the nature of forces that keep (O$_2$)$_4$ quadrates in
neighboring planes locked under high pressure, what is the role of
magnetic interactions -- all these questions are still open.
First-principles calculations \cite{Neaton:2002} demonstrate the
tendency of the O$_2$ molecules for dimerization and formation of
herringbone-type chains but failed to prove that the (O$_2$)$_4$
structure has the lowest energy.

With the account of these results, in the present paper we make an
attempt to elucidate the role of exchange interactions in
formation of nonmagnetic $\varepsilon$-phase and show how the
pressure-induced variation of the exchange constants may produce
strong distortion of crystal lattice.

\section{Intuitive considerations}
Analysis of the magnetic and structural properties of $\beta$-,
$\alpha$- and $\delta$-phases of solid oxygen shows
\cite{Freiman:2004} that exchange interactions in this crystal are
so strong that they are responsible not only for variation of
magnetic order but also produce rather noticeable deformations of
the crystal lattice. So, it seems reasonable to assume that
$\varepsilon$-phase makes no exclusion and its complicated and
surprising structure is mainly due to strong exchange
inter-molecule interaction that keeps quarters of O$_2$ molecules
as the independent chemical units, equalizes intermolecular
distances within these complexes and weakens intercluster bonds to
so extent that O$_8$ clusters can be approximately considered as
(magnetically) noninteracting units.

From general point of view, magnetic collapse (disappearance of
magnetic properties) observed in $\varepsilon$-phase may result
from coupling of 2, 3, or any other number of O$_2$ molecules in a
singlet spin state. The tendency of the O$_2$ molecules to form
such multimolecular clusters (consisting of 2, 3, 4 units) was
ascertained long ago in the optical spectra of $\alpha$-phase,
where two-, three- (at higher temperature) and four-molecule
dipole transition bands were directly observed and identified
\cite{Gaididei:1977}. Why, then, 4O$_2$ complex is more
favorable\footnote{~In the very recent paper
Ref.\onlinecite{Steudel:2007} the authors claim that according to
their quantum chemical calculations a ``rhomboid O$_8$ structure
of D$_2h$ symmetry is a stable species, that is, a local energy
minimum'' and point out that the ground state of a rhomboid should
be a singlet spin state. On the basis of density functional
calculations the authors also predict stability of the trimer
structure S$_6$ as corresponding to a local energy minimum on the
potential hypersurface.} than, say, dimer, 2O$_2$, or trimer,
3O$_2$?

\begin{figure}[htbp]
\begin{minipage}[b]{0.45\linewidth}
{\centerline{\includegraphics[width=1\textwidth]{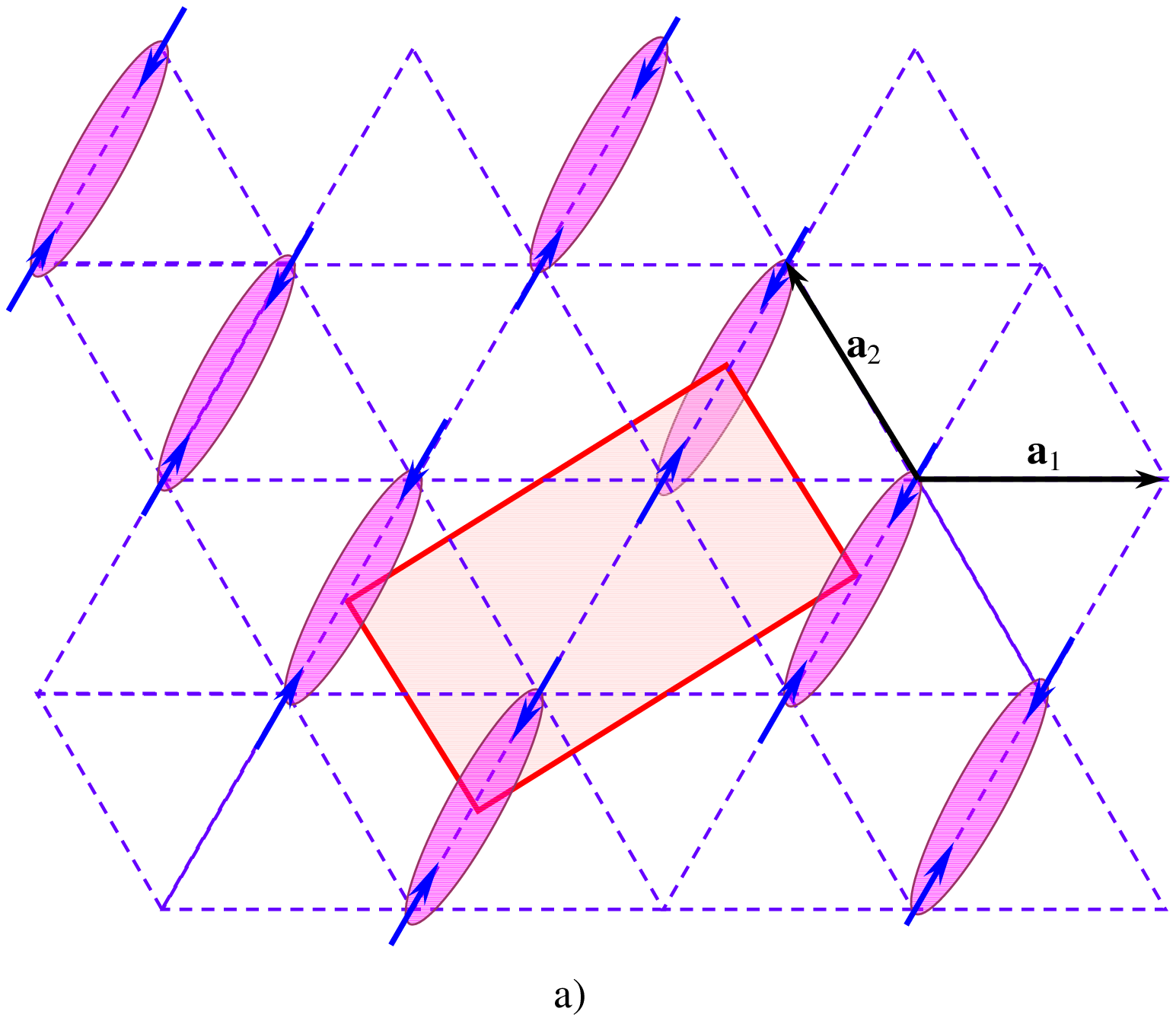}}
}\end{minipage}\hfill
\begin{minipage}[b]{0.45\linewidth}{\centerline{\includegraphics[width=1\textwidth]{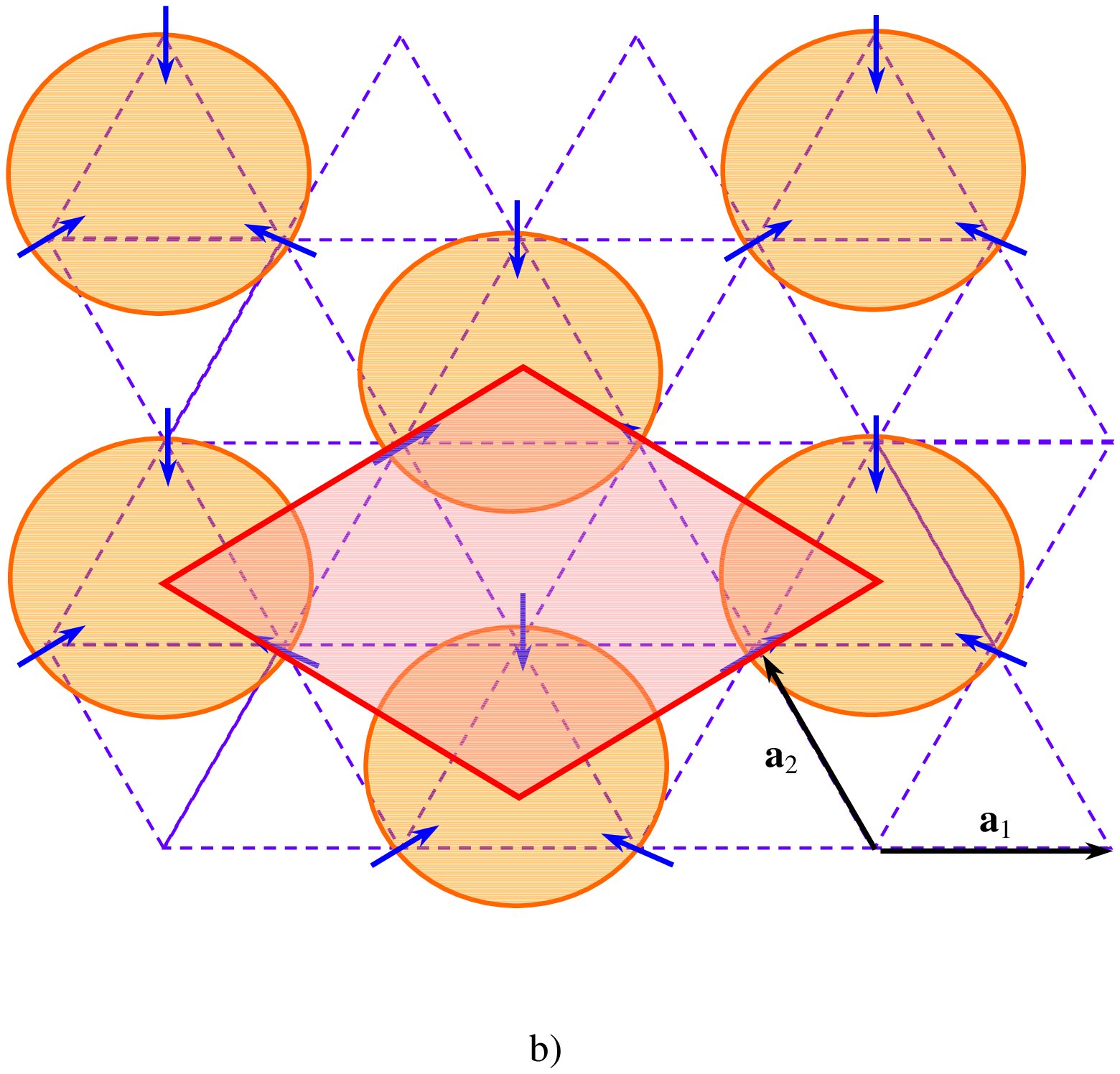}}}
\end{minipage}\caption{Covering of hexagonal plane with a) pairs
and b) triples of O$_2$ molecules. Arrows show directions of the
molecular shift in a corresponding optical mode, red parallelogram
is a unit cell of the
superstructure.\label{Fig_2-singlet}}\end{figure}

\begin{figure}[htbp]
{\centerline{\includegraphics[width=0.5\textwidth]{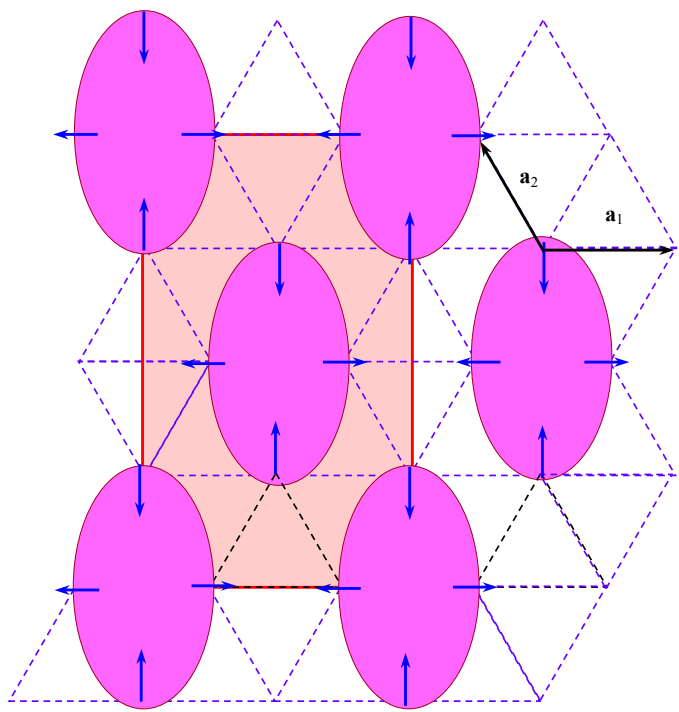}}
\caption{Covering of hexagonal plane with quarters (outlined with
ellipses) of O$_2$ molecules. Arrows show directions of the
molecular shift in a corresponding optical mode, red rectangle is
a unit cell of the
superstructure.\label{Fig_4-singlet}}}\end{figure}

 One of the possible reason for such a behavior is a weakness of van
 der Waals intermolecular forces in comparison with exchange
 interaction. As long as exchange interactions are not taken
 into account, O$_2$ molecules in $\alpha$-, $\beta$-, $\delta$-
 and $\varepsilon$-phases can be considered as noninteracting solid
 spheres packed in the most compact way, i.e. in a regular triangular lattice, within
the basal plane \cite{English:1974}(see also
Refs.\onlinecite{Cryocrystals_e, Freiman:2004}). Singlet complexes
concatenated by the exchange forces and decoupled from each other
may also be considered as noninteracting (or weakly interacting)
solids. Dimers themselves \cite{Gorelli:2001} are highly
anisotropic, formation of the decoupled pairs should produce
additional distortion of crystal lattice (see
Fig.\ref{Fig_2-singlet} a), so, O$_4$ complexes seem to be
unstable with respect to formation of herribone chains. In
contrary, 3O$_2$, 4O$_2$ and 7O$_2$ complexes may be invariant
with respect to rotation around 3-rd, 4-th, or, correspondingly,
6th order symmetry axis, and so are isotropic in the basal plane.
In turn, a hexagonal plane can be completed by the regular
triangulars (Fig. \ref{Fig_2-singlet}b), 60$^\circ$-angled
diamonds (Fig. \ref{Fig_4-singlet}), or hexagons that by
appropriate deformations may be transformed into highly symmetric
$n$-O$_2$ units. Lattice distortions shown in Figs.
\ref{Fig_2-singlet} and \ref{Fig_4-singlet} by arrows could be
classified (see Table\ref{irreducible representations}) according
to symmetry of different optical modes\footnote{~In contrast to
homogeneous deformations, an optical mode removes degeneracy of
the nearest neighbors intermolecular distance.}.


It is quite obvious that formation of trimers and sestets should
be accompanied by isotropic contraction of inter-atomic distances
within the complex, while formation of quaters (see below) is
related with anisotropic (shear) deformation of corresponding
rhombus. The second process seems to be energetically more
favorable because shear modes are usually much more soft compared
to isomorphic striction.

So, formation of quaters may be induced by increase of AFM
exchange coupling at high pressure and softening of appropriate
optical mode.

\begin{table}\label{irreducible representations}
  \centering
  \caption{Wave vectors and polarization of optical modes coupled with
  different singlet states. Representatives of stars $\mathbf{k}$ are
  given according to Kovalev's notations\cite{Kovalev:1986E}. Lattice vectors
$\mathbf{a}_1$, $\mathbf{a}_2$ and reciprocal lattice vectors
$\mathbf{b}_1$,  $\mathbf{b}_2$ are attributed to the hexagonal
pra-phase.}
  \begin{tabular}{|c|c|c|}
\hline Number of O$_2$ mol.&Wave vector&Polarization vector
\\\hline
2-dimer
&$\mathbf{k}_{12}=\mathbf{b}_1/2$&$\mathbf{a}_1+\mathbf{a}_2$
\\ \hline 3-trimer
&$\mathbf{k}_{13}=(\mathbf{b}_1+\mathbf{b}_2)/3$&$\sqrt{3}(\mathbf{a}_1+\mathbf{a}_2)+i(\mathbf{a}_1-\mathbf{a}_2)$\\
\hline 4-quater
&$\mathbf{k}_7=\mathbf{b}_1/2+\mathbf{b}_2/4$,&$u_x\sqrt{3}(\mathbf{a}_1+\mathbf{a}_2)+iu_y(\mathbf{a}_1-\mathbf{a}_2)$\\
&$\mathbf{b}_1/2-\mathbf{b}_2/4$&$iu_x\sqrt{3}(\mathbf{a}_1+\mathbf{a}_2)-u_y(\mathbf{a}_1-\mathbf{a}_2)$\\\hline
\end{tabular}
\end{table}

\section{Model}
\subsection{Order parameter}
Phenomenological description of $\varepsilon$-O$_2$ as a phase in
a series of
$\beta\rightarrow\alpha\rightarrow\delta\rightarrow\varepsilon$
transition is not so straightforward as of the other magnetic
phases. According to phase diagram \cite{Goncharenko:2005},
$\varepsilon$-phase may be obtained from both $\delta$- and
$\beta$-phases whose Brave lattices belong to the different space
groups, so, what phase should be considered as a parent phase?
Three-dimensional space group of $\varepsilon$-O$_2$ coincides
with that of $\alpha$-O$_2$, though in the phase diagram both
phases are separated with the high symmetry $\delta$-phase. What
is more, $\alpha$-, $\delta$- and $\varepsilon$-phases are
described by the same symmetry group within the basal plane. This
fact makes questionable the choice of the components of
deformation tensor as an order parameter of
$\delta\varepsilon$-transition.

The easiest way to overcome these difficulties is to accept that
all the magnetic phases, including $\varepsilon$-O$_2$, originate
from a virtual nonmagnetic phase viewed as a stack of regular
triangular planes. This assumption is based on the following
facts.
\renewcommand{\theenumi}{\roman{enumi}}
\renewcommand{\labelenumi}{\theenumi)}
\begin{enumerate}
  \item Crystal lattices of $\beta$-,
$\alpha$- and $\delta$-phases can be thought of as the different
modifications of the same hexagonal (space group 6/mmm) pra-phase
 in which the neighboring close-packed planes are shifted in [1100] direction \cite{gomo:2005}.
  \item Though the
O$_2$ lattice in $\varepsilon$-phase is strongly distorted within
the basal plane, as compared to lattice of $\alpha$-phase, an
angle between the bonds connecting the molecules in neighboring
O$_8$ clusters remains approximately equal to 60$^\circ$ in a wide
interval of pressures, as seen from the experiment
\cite{Fujihisa:2006}.
\end{enumerate}

From this point of view, the structural order parameter of
pra-phase -- $\varepsilon$-phase transition can be represented by
the amplitudes $u_x$, $u_y$  of the optical mode
\begin{equation}\label{optical_mode}
  u(\mathbf{n})=e^{i\mathbf{b}_1\mathbf{n}/2}\left[u_x\cos\frac{\mathbf{b}_2\mathbf{n}}{4}+u_y\sin\frac{\mathbf{b}_2\mathbf{n}}{4}\right],
\end{equation}
where vector $\mathbf{n}$ denotes position of a molecule within
the basal plane\footnote{~Order parameter of pra-to-$\beta$-,
$\alpha$- and $\delta$- phase transitions is coupled with a
transverse acoustic mode in [0001] direction.}.
 Macroscopic description of the magnetic state of $\varepsilon$-phase
 may be done with the use of spin-spin correlation functions.
 Discussion of this question is beyond the scope of the present
 paper.

It is interesting to note that the structural order parameter  in
the sequence of $\beta\rightarrow\alpha\rightarrow\delta$-phase
transitions (i.e. a function of mutual shift of the neighboring
close-packed planes in [1100] direction calculated with respect to
the initial non-shifted hexagonal stacking) is symmetry related to
the transverse acoustic modes propagating in [0001] and [1000]
directions (wave vectors parallel to $\mathbf{b}_3$ and
$\mathbf{b}_1$, correspondingly).

\subsection{Free energy and spin hamiltonian}
Different phases of solid oxygen and inter-phase transitions are
described on the basis of phenomenological expression for free
energy of the crystal. Substantial simplification of the model may
be achieved by neglection of interplane interactions. This
assumption is justified by noticeable difference in the variation
of in-plane and interplane distances in the course of
pressure-induced phase transitions.

As it was mentioned above, the magnetic and crystal structure of
$\alpha$- and $\delta$-phases is indistinguishable within the
$ab$-plane, so, herewith we consider the series of
$\beta\rightarrow\alpha\rightarrow\varepsilon$-transition. Gibbs
free energy $\Phi$ of the crystal is modeled as a function of
(two-dimensional) phonon amplitude $\mathbf{u}(\mathbf{k})$,
strain tensor components $u_{jk}$, invariant with respect to the
symmetry group of hexagonal pra-phase, plus magnetic contribution
into internal energy $E_{\rm mag}$:
\begin{eqnarray}\label{elastic+phonon}
  \Phi&=&\frac{1}{2}\sum_{j}K(\mathbf{k}_j)|\mathbf{u}(\mathbf{k}_j)|^2+\frac{c_{11}+c_{12}}{2}(u_{xx}+u_{yy})^2+\frac{c'}{2}[(u_{xx}-u_{yy})^2+4u_{xy}^2]+P(u_{xx}+u_{yy})\nonumber\\
&+&\sum_{j}\lambda^{({\rm iso})}_{\rm
  ph}(\mathbf{k}_{j})|\mathbf{u}(\mathbf{k}_{j})|^2(u_{xx}+u_{yy})+ \lambda^{({\rm an})}_{\rm
  ph}(\mathbf{k}_7)[u^2_x(\mathbf{k}_{7})-u^2_y(\mathbf{k}_{7})](u_{xx}-u_{yy})+E_{\rm
  mag}.
\end{eqnarray}
Vectors $\mathbf{k}_j$ in the above expression denote different
wave vectors, classified according to irreducible representations
of 6/mmm space group,  phenomenological constants
$K(\mathbf{k}_j)$ are proportional to the corresponding phonon
frequencies, coefficients $\lambda_{\rm
  ph}(\mathbf{k}_{j})$ originate from the crystal anharmonicity
  and describe nontrivial coupling between phonon amplitude and crystal lattice
  parameters. Last term in (\ref{elastic+phonon}) accounts for the
  (external) hydrostatic pressure $P$.

Magnetic contribution $E_{\rm mag}$ is calculated as an average of
spin-hamiltonian $\hat\mathcal{H}$ over a ground state
$|\Psi\rangle$ of the crystal: $E_{\rm mag}=\langle
\Psi|\hat\mathcal{H}|\Psi\rangle$, where
\begin{equation}\label{hamiltonian_general}
  \hat\mathcal{H}=\sum_{\mathbf{nm}}J(\mathbf{r}_{\mathbf{nm}})\hat\mathbf{S}_{\mathbf n}\hat\mathbf{S}_{\mathbf
  m},
\end{equation}
and summation is accomplished over the nearest and next to the
nearest neighbors separated by distance
$|\mathbf{r}_{\mathbf{nm}}|$. Magnetoelastic part of the internal
energy is derived from the expression (\ref{hamiltonian_general})
with due account of space dependence of the exchange integral
$J(\mathbf{r}_{\mathbf{nm}})$.

Once the ground state of magnetic subsystem is calculated,
structure and stability conditions of a phase can be determined by
minimization of free energy (\ref{elastic+phonon}) with respect to
phonon amplitudes and deformation tensor components.

As it was already mentioned, $\beta$- and $\alpha$-phases posses a
kind of  the N\'eel spin ordering, that can be described
macroscopically by assigning an average value $\langle
\mathbf{S}_\mathbf{n}\rangle$  to the spin vector at each site.
\begin{figure}[htbp] \centering{
\epsfig{file=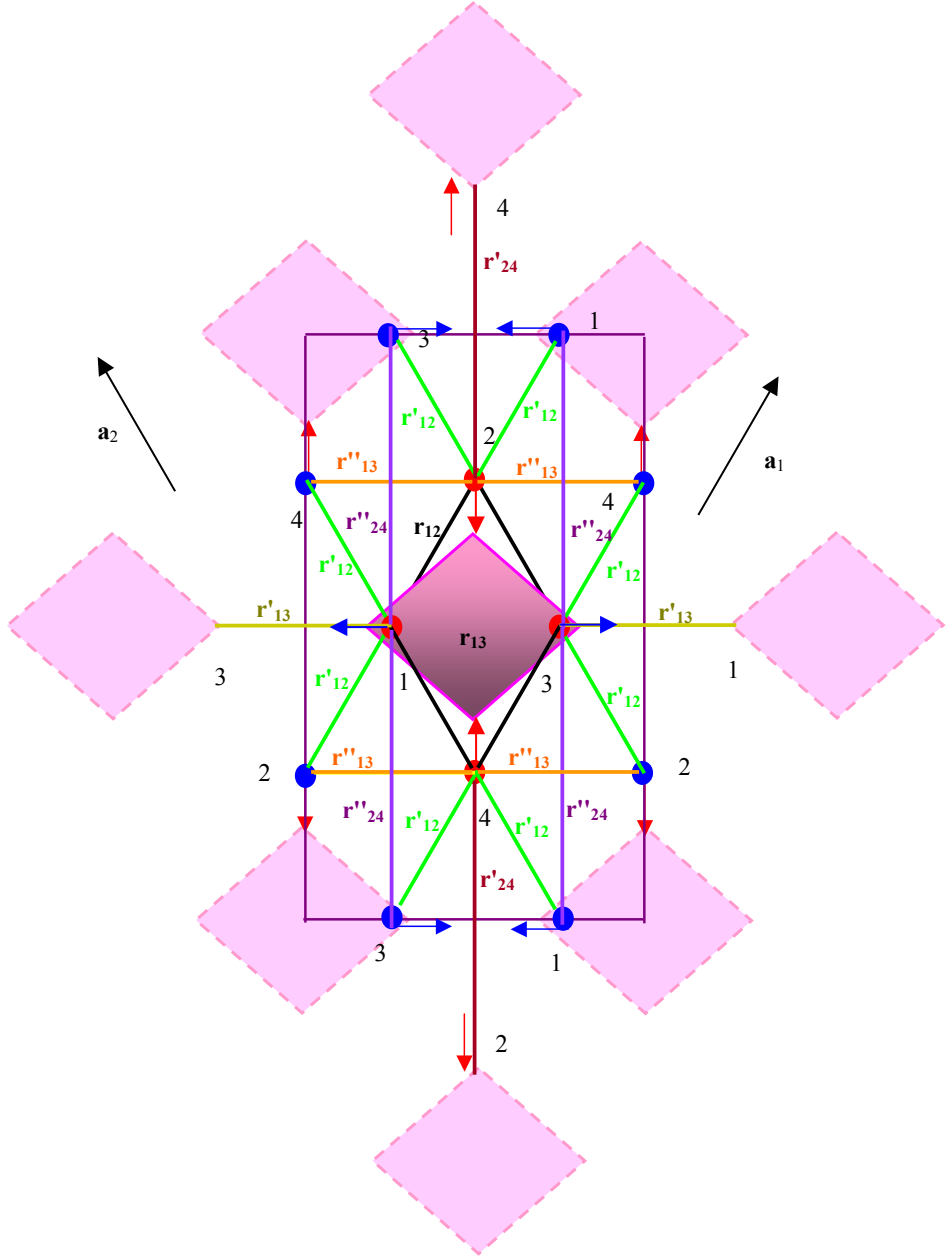,width=0.7\textwidth}} \caption{Primitive
cell of $\varepsilon$-phase. Magnetically coupled quaters of
molecules are outlined by quadrates. Red and blue arrows show the
shift of O$_2$ molecules in the course of phase transition.
Vectors $\mathbf{a}_{1,2}$ are the Brave lattice vectors of the
hexagonal pra-phase. }\label{Fig_epsilon}\end{figure}

A ground state $|\Psi_\varepsilon\rangle$ of $\varepsilon$-phase
is a true eigen function of spin-hamiltonian
(\ref{hamiltonian_general}) and is calculated within an assumption
of magnetically decoupled O$_8$ clusters. In other words,
$|\Psi_\varepsilon\rangle$ may be represented as unentangled
combination of quaters wave functions $|\psi_\mathbf{n}\rangle$,
that satisfies equation
\begin{equation}\label{eigen_equation}
\hat\mathcal{H}^{\rm (intra)}|\Psi_\varepsilon\rangle=E^{\rm
(intra)} |\Psi_\varepsilon\rangle, \quad
|\Psi_\varepsilon\rangle\equiv\prod_{\mathbf{n}}|\psi_\mathbf{n}\rangle,
\end{equation}
with the hamiltonian of intra-cluster interactions written as
follows
\begin{eqnarray}\label{intra}
\hat\mathcal{H}^{\rm (intra)}&=&\sum_{\mathbf{n}}\left\{J(\mathbf{r}_{12})[\hat\mathbf{S}_{1\mathbf{n}}\hat\mathbf{S}_{2\mathbf{n}}+\hat\mathbf{S}_{2\mathbf{n}}\hat\mathbf{S}_{3\mathbf{n}}+\hat\mathbf{S}_{3\mathbf{n}}\hat\mathbf{S}_{4\mathbf{n}}+\hat\mathbf{S}_{4\mathbf{n}}\hat\mathbf{S}_{1\mathbf{n}}]\right.\nonumber\\
&+&\left.
J(\mathbf{r}_{13})\hat\mathbf{S}_{1\mathbf{n}}\hat\mathbf{S}_{3\mathbf{n}}+J(\mathbf{r}_{24})\hat\mathbf{S}_{2\mathbf{n}}\hat\mathbf{S}_{4\mathbf{n}}\right\}.
\end{eqnarray}
Here the vectors $\mathbf{n}$ define the positions of O$_8$
cluster in a superstructure with lattice vectors
$\mathbf{a}'_1=2\mathbf{a}_1$ and $\mathbf{a}'_2=2\mathbf{a}_2$.
For the sake of simplicity we use the rectangular ``unit cell''
which contains two clusters (see Fig.\ref{Fig_epsilon}). Choice of
the unit cell corresponds to one of three different domains of
$\alpha$-phase. The positions of the individual O$_2$ molecules
(labelled with number 1, 2, 3, 4) within the cluster are defined
with the basis vectors $\pm\mathbf{\tau}_1\equiv \pm\mathbf{a}_1$
and $\pm\mathbf{\tau}_2\equiv \pm \mathbf{a}_2$.

 Interaction between the clusters with account of the next-to
 nearest neighbors is described by operator $\mathcal{H}^{\rm
(inter)}$ (see Fig.\ref{Fig_epsilon} for notations)
\begin{eqnarray}\label{inter}
&&\mathcal{H}^{\rm
(inter)}=\frac{1}{2}\sum_{\mathbf{n}}\left\{J(\mathbf{r}'_{12})[\hat\mathbf{S}_{1\mathbf{n}}(\hat\mathbf{S}_{2(\mathbf{n}-\mathbf{\tau}_1)}+\hat\mathbf{S}_{4(\mathbf{n}+\mathbf{\tau}_2)})+\hat\mathbf{S}_{3\mathbf{n}}(\hat\mathbf{S}_{2(\mathbf{n}-\mathbf{\tau}_2)}+\hat\mathbf{S}_{4(\mathbf{n}+\mathbf{\tau}_1)})\right.\nonumber\\
&&+\hat\mathbf{S}_{2\mathbf{n}}(\hat\mathbf{S}_{1(\mathbf{n}+\mathbf{\tau}_1)}+\hat\mathbf{S}_{3(\mathbf{n}+\mathbf{\tau}_2)})+\hat\mathbf{S}_{4\mathbf{n}}(\hat\mathbf{S}_{1(\mathbf{n}-\mathbf{\tau}_2)}+\hat\mathbf{S}_{3(\mathbf{n}-\mathbf{\tau}_1)})]\nonumber\\
&&+J(\mathbf{r}'_{13})(\hat\mathbf{S}_{1\mathbf{n}}\hat\mathbf{S}_{3(\mathbf{n}+\mathbf{\tau}_1-\mathbf{\tau}_2)}+\hat\mathbf{S}_{3\mathbf{n}}\hat\mathbf{S}_{1(\mathbf{n}-\mathbf{\tau}_1+\mathbf{\tau}_2)})\nonumber\\
&&+J(\mathbf{r}''_{24})\left[\hat\mathbf{S}_{1\mathbf{n}}(\hat\mathbf{S}_{3(\mathbf{n}+\mathbf{\tau}_1+\mathbf{\tau}_2)}+\hat\mathbf{S}_{3(\mathbf{n}-\mathbf{\tau}_1-\mathbf{\tau}_2)})
+\hat\mathbf{S}_{3\mathbf{n}}(\hat\mathbf{S}_{1(\mathbf{n}+\mathbf{\tau}_1+\mathbf{\tau}_2)}+\hat\mathbf{S}_{1(\mathbf{n}-\mathbf{\tau}_1-\mathbf{\tau}_2)})\right]\nonumber\\
&&+J(\mathbf{r}''_{13})\left[\hat\mathbf{S}_{2\mathbf{n}}(\hat\mathbf{S}_{4(\mathbf{n}+\mathbf{\tau}_1-\mathbf{\tau}_2)}
+\hat\mathbf{S}_{4(\mathbf{n}-\mathbf{\tau}_1+\mathbf{\tau}_2)})+\hat\mathbf{S}_{4\mathbf{n}}(\hat\mathbf{S}_{2(\mathbf{n}+\mathbf{\tau}_1-\mathbf{\tau}_2)}+\hat\mathbf{S}_{2(\mathbf{n}-\mathbf{\tau}_1+\mathbf{\tau}_2)})\right]\nonumber\\
&&+\left.J(\mathbf{r}'_{24})[\hat\mathbf{S}_{2\mathbf{n}}\hat\mathbf{S}_{4(\mathbf{n}+\mathbf{\tau}_1+\mathbf{\tau}_2)}+\hat\mathbf{S}_{4\mathbf{n}}\hat\mathbf{S}_{2(\mathbf{n}-\mathbf{\tau}_1-\mathbf{\tau}_2)})]\right\},
\end{eqnarray}
so that spin-hamiltonian (\ref{hamiltonian_general}) is
represented as a  sum:
\begin{equation}\label{full_ham}
  \hat\mathcal{H}=\hat\mathcal{H}^{\rm (intra)}+\hat\mathcal{H}^{\rm
(inter)}.
\end{equation}
In the $\varepsilon$-phase the first term in (\ref{full_ham}) is
responsible for formation of the ground state, while the second
one describes contribution that arises from excitations. In
$\alpha$-, $\delta$ and $\beta$-phases both terms  contribute
equally into magnetic energy of the crystal.
\section{$\varepsilon$-phase}
\subsection{Magnetic structure}
It was already mentioned that according to experimental data, the
units O$_8$ form a common singlet state, while each O$_2$ molecule
possesses spin $S_j=1$, $j=1-4$. So, it is convenient to express
spin-state $\psi$ of O$_8$ cluster in terms of the eigen functions
$|0\rangle$, $|\pm 1\rangle$ of spin operators $\hat{S}_{j}^{Z}$,
where $Z$ is a quantization axis.

According to general theorem of quantum mechanics, singlet state
in such a system has 3 representations (among 81 basic vectors)
with the spin wave functions that could be easily found from the
conditions
\begin{equation}\label{singlet_equation}
 \left( \sum_{j=1}^4\hat\mathbf{S}_j\right)^2\psi_{\rm singlet}=0,
 \quad  \sum_{j=1}^4\hat{S}_{j}^{Z}\psi_{\rm singlet}=0.
\end{equation}
Obviously, $\psi_{\rm singlet}$ is also an eigen function of
hamiltonian $\hat\mathcal{H}^{\rm (intra)}$.

Additional simplification of the problem may be achieved by
account of permutation symmetry group. All three singlet states
should have different symmetry with respect to permutations of
molecules within the cluster and hence, correspond to the
different eigen values of operators (see Table \ref{table_1})
\begin{equation}\hat P_1\equiv
(\hat\mathbf{S}_1,\hat\mathbf{S}_2)+(\hat\mathbf{S}_2,\hat\mathbf{S}_3)+(\hat\mathbf{S}_3,\hat\mathbf{S}_4)+(\hat\mathbf{S}_4,\hat\mathbf{S}_1)$$
$$\hat P_2\equiv(\hat\mathbf{S}_1,\hat\mathbf{S}_3), \quad \hat
P_3\equiv(\hat\mathbf{S}_2,\hat\mathbf{S}_4)\end{equation}

Finally, the singlet wave functions may be written in the
following form:
\begin{eqnarray}\label{wave_function singlet_1}
  &&\psi_{\rm gr}^{\rm (singlet)}=\frac{1}{\sqrt{5}}(|1\overline{1}1\overline{1}\rangle+|\overline{1}1\overline{1}1\rangle)+\frac{1}{3\sqrt{5}}\left[2|0000\rangle+|010\overline{1}\rangle+|10\overline{1}0\rangle\right.\nonumber\\
  &&\left.|0\overline{1}01\rangle+|\overline{1}010\rangle-{3\over
  2}(|001\overline{1}\rangle+|01\overline{1}0\rangle+|1\overline{1}00\rangle+|\overline{1}001\rangle+|00\overline{1}1\rangle+|0\overline{1}10\rangle\right.\nonumber\\
&&+\left. |\overline{1}100\rangle+|100\overline{1}\rangle)+{1\over
2}(|11\overline{1}\overline{1}\rangle+|\overline{1}\overline{1}11\rangle+|1\overline{1}\overline{1}1\rangle+|\overline{1}11\overline{1}\rangle)
  \right].
\end{eqnarray}
\begin{eqnarray}\label{wave_function singlet_2}
  &&\psi_{\rm ex 1}^{\rm
  (singlet)}=\frac{1}{2\sqrt{3}}\left[|1\overline{1}00\rangle+|001\overline{1}\rangle+|\overline{1}100\rangle+|00\overline{1}1\rangle+
 |11\overline{1}\overline{1}\rangle+|\overline{1}\overline{1}11\rangle\right.\nonumber\\
  &&\left.-|1\overline{1}\overline{1}1\rangle-|\overline{1}11\overline{1}\rangle-|100\overline{1}\rangle-|01\overline{1}0\rangle-|\overline{1}001\rangle-|0\overline{1}10\rangle\right].
\end{eqnarray}
\begin{eqnarray}\label{wave_function singlet_3}
  &&\psi_{\rm ex 2}^{\rm
  (singlet)}=\frac{1}{3}\left[|11\overline{1}\overline{1}\rangle+|\overline{1}\overline{1}11\rangle+|1\overline{1}\overline{1}1\rangle+|\overline{1}11\overline{1}\rangle+|0000\rangle\right.\nonumber\\
   &&\left.-|010\overline{1}\rangle-|0\overline{1}01\rangle-|10\overline{1}0\rangle-|\overline{1}010\rangle\right].
\end{eqnarray}

\begin{table}
  \centering
  \caption{Eigen values of operators $\hat P_1$, $\hat P_2$, $\hat P_3$ in a singlet subspace, cluster energy $\langle\hat \mathcal{H}^{\rm (intra)}\rangle/N$ per molecule for arbitrary intermolecular spacing, equilibrium angle $\varphi$ between intermolecular bonds within the cluster and corresponding equilibrium energy $E_\varepsilon$ per molecule.}\label{eigen_values}
  \begin{tabular}{|c|c|c|c|c|c|}
\hline
    Function & $\hat P_1$ & $\hat P_{2,3}$ &$\langle\hat \mathcal{H}^{\rm (intra)}\rangle/N$&$\varphi_{\rm eq}$&$E_\varepsilon$\\\hline
    $\psi_{\rm gr}^{\rm (singlet)}$ & -6 & 1 &$[J(\mathbf{r}_{13})+J(\mathbf{r}_{24})-6J(\mathbf{r}_{12})]/4$&$\pi/4$&$[J(\sqrt{2}a)-3J(a)]/2$\\\hline
    $\psi_{\rm ex 1}^{\rm
  (singlet)}$ & -4 & 0 &$-J(\mathbf{r}_{12})$&arb.&$-J(a)$\\\hline
    $\psi_{\rm ex 2}^{\rm
  (singlet)}$ & 0 & -2 &$-[J(\mathbf{r}_{13})+J(\mathbf{r}_{24})]/2$&$\pi/4$&$-J(\sqrt{2}a)$\\ \hline
  \end{tabular}\label{table_1}
\end{table}
In order to find out what of three functions (\ref{wave_function
singlet_1})-(\ref{wave_function singlet_3}) describes the ground
state of hamiltonian (\ref{full_ham}), we compare corresponding
eigen values (see Table\ref{table_1}, the 4th column). Taking into
account antiferromagnetic character of the exchange interaction
($J(\mathbf{r})>0$), the fact, that $J(\mathbf{r})$ monotonically
decreases with intermolecular distance $r$, and geometrical
relation $r_{12}\le r_{13}<r_{24}$, one can easily verify that
\begin{equation}\label{comparison}
  \langle\hat \mathcal{H}^{\rm (intra)}\rangle_{\rm gr}<\langle\hat \mathcal{H}^{\rm (intra)}\rangle_{\rm ex 1}<\langle\hat \mathcal{H}^{\rm (intra)}\rangle_{\rm ex 2},
\end{equation}
and the required ground state is $\psi_{\rm gr}^{\rm
  (singlet)}$ (see eq.(\ref{wave_function singlet_1})). We have
  also implicitly taken into account an obvious fact that an average value of $\langle\hat \mathcal{H}^{\rm
  (inter)}\rangle$ in any singlet state is exactly zero.

  It is interesting to compare $\psi_{\rm gr}^{\rm (singlet)}$ with the AFM N\'eel state observed in $\alpha$-phase,
  where all the nearest neighbors are coupled antiferromagnetically.
  In terms of O$_2$ spin states it means, that the most preferable
 combinations  are $|1\overline{1}1\overline{1}\rangle$ and
 $|\overline{1}1\overline{1}1\rangle$. States $\psi^{(\rm ex 1,2)}_{\rm
  singlet}$ are orthogonal to a subspace spanned over the ``N\'eel-state'' vectors $|1\overline{1}1\overline{1}\rangle$ and
 $|\overline{1}1\overline{1}1\rangle$. In contrary, $\psi_{\rm gr}^{\rm (singlet)}$ belongs to this subspace with 0.4 probability.

We may also compare eigen values of hamiltonian (\ref{full_ham})
in a singlet state for different clusters: hypothetical dimer,
trimer and already described quater. It is obvious that for 2O$_2$
and 3O$_2$ complexes the singlet state is kept by the nearest
neighbor interactions only, corresponding eigen values of
hamiltonian (that could be found without explicit expression for
wave function) are
\begin{eqnarray}\label{energy_dimer}
\langle\hat \mathcal{H}^{\rm (intra)}\rangle_{\rm
dim}&=&-NJ(\mathbf{r}_{12}), \quad {\textrm {for \,
dimer}};\nonumber\\ \langle\hat \mathcal{H}^{\rm
(intra)}\rangle_{\rm
trim}&=&-N\left[\frac{2}{3}J(\mathbf{r}_{12})+\frac{1}{3}J(\mathbf{r}_{13})\right],\quad
{\textrm {for \, trimer}}.\end{eqnarray} where $N$ is the number
of O$_2$ molecules. In the case of monotonically decreasing AFM
exchange and fixed $r_{12}\le r_{13}<r_{24}$ values
\begin{equation}\label{comparison_2}
  \langle\hat \mathcal{H}^{\rm (intra)}\rangle_{\rm gr}^{\rm
  (singlet)}<\langle\hat \mathcal{H}^{\rm (intra)}\rangle_{\rm
dim}=\langle\hat \mathcal{H}^{\rm (intra)}\rangle_{\rm ex 1}^{\rm
  (singlet)}\le\langle\hat
\mathcal{H}^{\rm (intra)}\rangle_{\rm trim}<\langle\hat
\mathcal{H}^{\rm (intra)}\rangle_{\rm ex 2}^{\rm
  (singlet)}.
\end{equation}
So, the magnetic energy of the crystal in the state with $S=0$
 takes on its minimum value when the number of O$_2$ molecules in
 a singlet group is at least $n=$4.


\subsection{Distortion of crystal lattice}
It was already mentioned that magnetic interactions in solid
oxygen are so strong that they cause large distortion of crystal
lattice. This effect was observed, e.g., in the course of
$\alpha\beta$-transitions where in-plane lattice deformation
achieved nearly 5\% \cite{Krupskii:1979E}. In the
$\varepsilon$-phase the effect of magnetoelastic interactions is
even more pronounced, though very unusual, because in this case
lattice distortion is produced by magnetic collapse, not by
magnetic ordering.

\begin{figure}[htbp] \centering{
\epsfig{file=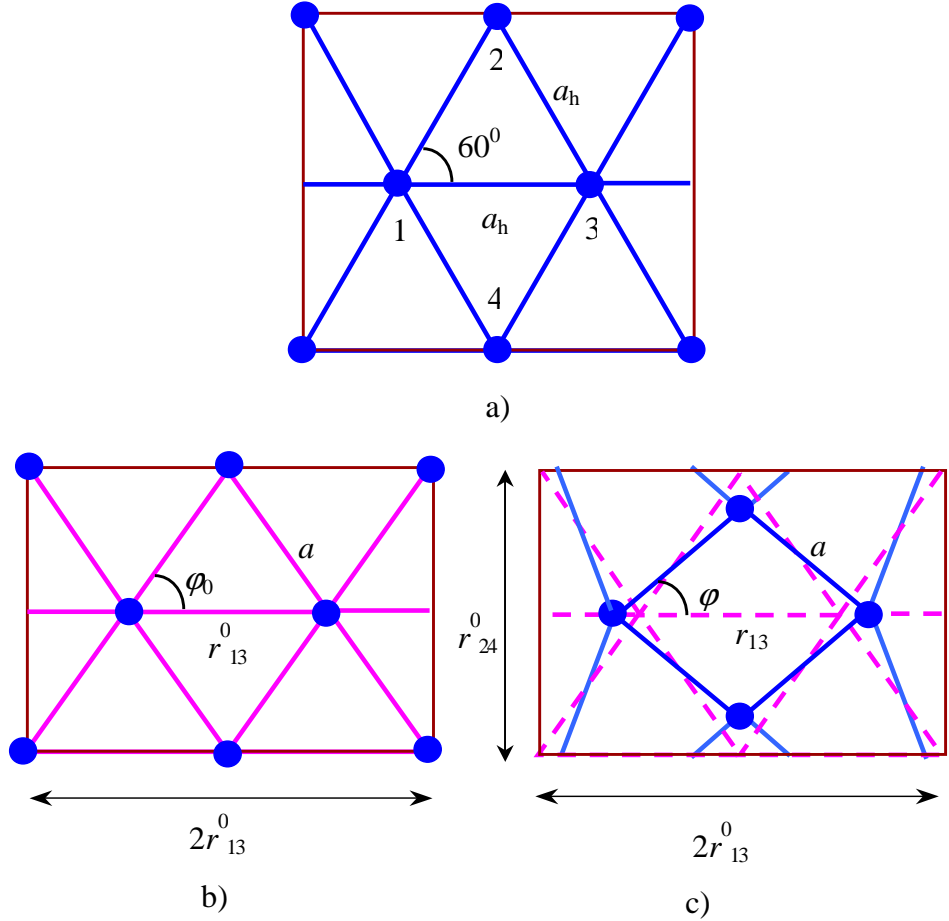,width=0.7\textwidth}} \caption{Two-step
distortion of the crystal lattice of hexagonal pra-phase (a):
\emph{i}) homogeneous deformation (b); \emph{ii}) inhomogeneous
distortion of rhombus (c).}\label{Fig_epsilon_def}\end{figure}

In-plane structure of $\varepsilon$-phase can be considered as a
result of  two-step distortion\footnote{~In approximation of
linear elasticity the sequence of steps is immaterial. An accepted
sequence is convenient from methodological point of view.} of the
ideal hexagonal basal plane of pra-phase ($\beta$-) with the
lattice constant $a_h$ (see Fig.\ref{Fig_epsilon_def}): \emph{i})
homogeneous deformation which changes scales in $X$ and $Y$
direction:
\begin{equation}\label{scale}
r_{13}^{(0)}=a_h(1+u_{xx}), \quad
r_{24}^{(0)}=\sqrt{3}a_h(1+u_{yy});
\end{equation}
and \emph{ii}) inhomogeneous distortion of rhombus formed by the
in-cluster molecules 1, 2, 3, 4. In fact, this means that the
virtual intermediate state (after step \emph{i})) has an
$\alpha$-type lattice.

 Symmetry condition that
the distances $r_{jk}$  between the pairs of molecules $jk=12, 23,
34$, and 41 within the cluster are equal, $r_{jk}=a$, makes it
possible to introduce very convenient and obvious parametrization
using an angle $\varphi$ between the directions to nearest
neighbors: $r_{13}=2a\cos\varphi$, $r_{24}=2a\sin\varphi$, and
\begin{equation}\label{parametrization}
u_x=a\cos\varphi-\frac{1}{2}r_{13}^{(0)}\equiv
a(\cos\varphi-\cos\varphi_0),\quad
u_y=\frac{1}{2}r_{24}^{(0)}-a\sin\varphi=a(\sin\varphi_0-\sin\varphi),
\end{equation}
where $r_{13}^{(0)}=2a\cos\varphi_0$,
$r_{24}^{(0)}=2a\sin\varphi_0$ are interatomic distances in the
reference frame.

 The positions of O$_2$ molecules in
$\varepsilon$-phase are then calculated by minimization of Gibbs'
free energy (\ref{elastic+phonon}) with respect to the components
of deformation tensor $u_{jk}$ and angle $\varphi$. Last term,
$E_{\rm mag}$, is the magnetic energy in the singlet ground state,
\begin{equation}\label{mag_energy_epsilon}
  E_{\rm mag}=\frac{N}{4}\left[J(\mathbf{r}_{13})+J(\mathbf{r}_{24})-6J(\mathbf{r}_{12})\right].
\end{equation}

Thus, expression (\ref{elastic+phonon}) can be rewritten as
\begin{eqnarray}\label{elastic+phonon_epsilon}
  &\Phi&=\frac{1}{2}(c_{11}+c_{12})(u_{xx}+u_{yy})^2+\frac{1}{2}c'[(u_{xx}-u_{yy})^2+4u_{xy}^2]+P(u_{xx}+u_{yy})\nonumber\\
  &+&2a^2\sin^2\frac{\varphi-\varphi_0}{2}\left[K(\mathbf{k}_7)+\lambda^{({\rm iso})}_{\rm
  ph}(\mathbf{k}_7)(u_{xx}+u_{yy})-\lambda^{({\rm an})}_{\rm
  ph}(\mathbf{k}_7)\cos(\varphi+\varphi_0)(u_{xx}-u_{yy})\right]
  \nonumber\\
&+&\frac{N}{4}\left[J(2a\sin\varphi)+J(2a\cos\varphi)-6J(a)\right].
\end{eqnarray}

Minimization conditions ($\partial \Phi/\partial \xi_j=0$,
$|\partial^2 \Phi/\partial \xi_j\partial \xi_k|>0$,
$\xi_j=\varphi, u_{xx}, u_{yy}$) give rise to the following
equation for $\varphi$:
 \begin{equation}\label{equilibrium_angle}
\frac{dJ(r_{13})}{dr}\cos\varphi+\frac{dJ(r_{24})}{dr}\sin\varphi+\frac{4K(\mathbf{k}_7)a}{N}\sin(\varphi-\varphi_0)=0\nonumber\\
\end{equation}
Assuming the softening of the optical mode $\mathbf{k_7}$ in the
vicinity of phase transition point, so that $K(\mathbf{k}_7)a^2\ll
a|dJ(r)/dr|N$, we can neglect last term in
(\ref{equilibrium_angle}). Then, equation
(\ref{equilibrium_angle}) has an obvious solution $\varphi_{\rm
eq}=\pi/4$ (and automatically, $r_{13}=r_{24}$). This means that
the four molecules in the ground singlet state are situated in the
corners of quadrate and it is \emph{the exchange interaction
within the cluster} that keeps the molecules in that state. Such a
symmetric arrangement of molecules seems to be quite natural in
the case when the exchange forces are the strongest interactions
in the system. Really, in the ground state (\ref{wave_function
singlet_1}) the molecules in the neighboring corners (12, 23, 34,
and 41) with high probability have opposite spins and thus are
attracted to each other, due to antiferromagnetic character of
exchange forces. In the contrary, the molecules in the opposite
corners (pairs 13 and 24) have parallel spins and are therefore
repulsed. The energy of repulsion is minimized when the average
distant between corresponding molecules is as much as possible,
this can be achieved in a symmetrical combination like quadrate.
Small deflection (e.g., 96$^\circ$ and 84$^\circ$ at 17.6 GPa)
from the right angle observed in the experiment
\cite{Lundegaard:2006} may be calculated from
(\ref{equilibrium_angle}) with account of contribution from the
optical mode:
\begin{equation}\label{equilibrium_angle_correct}
  \varphi=\frac{\pi}{4}-\frac{4\sqrt{2}K(\mathbf{k}_7)a\sin(\pi/4-\varphi_0)}{NJ'(r_{13})}.
\end{equation}
According to Ref.\onlinecite{Fujihisa:2006},
$\varphi_0\equiv\arctan(r^{(0)}_{24}/r^{(0)}_{13})=53^\circ$.
Using the most elaborated phenomenological form \cite{Etters:1985,
Etters:1988} of space dependence for
\begin{equation}\label{space-dependence}
  J(r)=J_0\exp[-\alpha(r-r_0)+\beta(r-r_0)^2],\quad 2.6\le r\le
  4.2,
\end{equation}
with $J_0=60$~K, $\alpha=3.5$~\AA$^{-1}$, $\beta=1.2$~\AA$^{-2}$,
$r_0=3.1854$~\AA, and taking $a=2.18$\AA
\cite{Lundegaard:2006,Fujihisa:2006}, we get an upper limit for
$K(\mathbf{k}_7)/N\le 9.2$~K/\AA$^{2}$, while estimated value of
$J'(r_{13})/a\ge 80$~K/\AA$^{2}$.
Considering $K$ as a stiffness constant of intermolecular bonds,
one obtains the characteristic frequency 16.2~cm$^{-1}$ which is
much smaller than the frequencies of optical modes ($\ge
300$~cm$^{-1}$) calculated in Ref.\onlinecite{Neaton:2002} and the
frequency 1,38~cm$^{-1}$ of Raman mode corresponding to the
antisymmetric stretching motion of the four O$_2$ molecules,
coupled in diagonal pairs \cite{Lundegaard:2006}.

 Stability condition of the ``quadrate'' solution
\begin{equation}\label{stability_phi}
  \frac{d^2J(r)}{dr^2}+\frac{2K(\mathbf{k}_7)}{N}\cos(\pi/4-\varphi_0)>0
\end{equation}
 is obviously satisfied, because according to Ref.\onlinecite{Freiman:2004} $J(r)$ is a monotonically decreasing concave function (see e.g. (\ref{space-dependence})) of
intermolecular distance, and $K(\mathbf{k}_7)>0$ (from the
condition of crystal lattice stability).

So, even in magnetically neutral state the exchange interactions
play a role of a motive force that changes crucially an angle
$\varphi$ between intermolecular bonds.

Analysis of the expression (\ref{elastic+phonon_epsilon}) makes it
possible to calculate shear deformation $u_{xx}-u_{yy}$ and
isotropic striction $u_{xx}+u_{yy}$ within the plane:
\begin{eqnarray}\label{equlibrium}
&&u_{xx}-u_{yy}=\frac{2a^2\lambda^{({\rm an})}_{\rm
  ph}(\mathbf{k}_7)}{c'}\cos(\pi/4+\varphi_0),\nonumber\\
&&u_{xx}+u_{yy}=-\frac{1}{c_{11}+c_{12}}[P+2\lambda^{({\rm
iso})}_{\rm
  ph}(\mathbf{k}_7)a^2\sin^2\frac{\pi/4-\varphi_0}{2}]
\end{eqnarray}
Space dependence of the exchange constant $J(r)$ does not
contribute into macroscopic deformation, because O$_8$ clusters
are supposed to be decoupled from each other. So, shear
deformation of $\varepsilon$-phase is due solely to anharmonicity
(coupling constant $\lambda^{({\rm an})}_{\rm
  ph}(\mathbf{k}_7)$) of crystal lattice. Isotropic striction
  $u_{xx}+u_{yy}$ describes relative change of the in-plane square.
  From the pressure dependence of lattice parameters
  \cite{Lundegaard:2006} we can estimate the in-plane compressive modulus
$c_{11}+c_{12}=88$~GPa. From the value of jump of isotropic
striction in the $\delta\varepsilon$-transition point, $\Delta
(u_{xx}+u_{yy})=0.019$ we estimate isotropic anharmonicity
constant $\lambda^{({\rm iso})}_{\rm
  ph}(\mathbf{k}_7)=1.1\cdot 10^5$~K/\AA$^2$.

\section{Comparison with $\alpha$- and $\delta$-phases}
In the previous section is was shown that once the singlet ground
state is formed, crystal lattice should be distorted in a
described manner, due to strong exchange interactions and reduced
optical phonon frequency. But what about the inverse mechanism,
can the crystal lattice of $\delta$-($\alpha$-phase) be unstable
with respect to $u(\mathbf{k}_7)$ distortions?

To answer this question, we minimize Gibb's potential
(\ref{elastic+phonon}) assuming the presence of the collinear
long-range AFM order. In this case $\langle\hat S^2_n\rangle=2$,
$\langle\hat S_{n}^{Z}\rangle=\pm
 1$ at a site $\mathbf{R}_\mathbf{n}$, and spin polarization alters from 1 to -1 when shifted through the vectors $\mathbf{a}_1$, $\mathbf{a}_2$.
 Taking into account the locations of molecules 1-4 (see
 Fig.\ref{Fig_epsilon}), it can be easily seen that
\begin{equation}\label{average_alpha}
 \langle\hat \mathbf{S}_1\hat \mathbf{S}_2\rangle=\langle\hat
\mathbf{S}_2\hat
 \mathbf{S}_3\rangle=\langle\hat \mathbf{S}_3\hat
 \mathbf{S}_4\rangle=\langle\hat \mathbf{S}_4\hat
 \mathbf{S}_1\rangle=-1,\quad \langle\hat \mathbf{S}_1\hat
 \mathbf{S}_3\rangle=\langle\hat \mathbf{S}_2\hat
 \mathbf{S}_4\rangle=1
\end{equation}
 Substituting these values into (\ref{full_ham}) we obtain
 expression for the magnetic energy $E_{AFM}$ of AFM state:
\begin{eqnarray}\label{mag_energy_AF}
  E_{\rm mag}&\equiv& E_{AFM}=E^{\rm cluster}_{AFM}+E^{\rm int}_{AFM}=\frac{N}{4}\left[J(r_{13})+J(r_{24})-4J(r_{12})\right]\nonumber\\
  &+& \frac{N}{4}\left[ J(r'_{13})+J(r'_{24})+2(J(r''_{13})+J(r''_{24}))-4J(r'_{12})\right],
\end{eqnarray}
where the first term describes interactions inside the cluster and
the second one is responsible for interaction energy.

Using the same parametrization of shift components $u_x, u _y$ as
in (\ref{parametrization}) we express all the intermolecular
distances (see Fig.\ref{Fig_epsilon}) in the expression
(\ref{mag_energy_AF})
 in terms of $\varphi$, $a$ as follows:
\begin{eqnarray}\label{distances_2}
r'_{12}&=&a\sqrt{(2\cos\varphi_0-\cos\varphi)^2+(2\sin\varphi_0-\sin\varphi)^2},\\
 r'_{13}&=&2a(2\cos\varphi_0-\cos\varphi),\quad
 r''_{13}=2a\sqrt{\cos^2\varphi_0+(\sin\varphi_0-\sin\varphi)^2},\nonumber\\
r'_{24}&=&2a(2\sin\varphi_0-\sin\varphi),\quad
r''_{24}=2a\sqrt{\sin^2\varphi_0+(\cos\varphi_0-\cos\varphi)^2}.\nonumber
\end{eqnarray}
 Analysis of the expressions (\ref{elastic+phonon}), (\ref{mag_energy_AF}) shows that the conditions of minimum $\partial
\Phi/\partial \varphi=0$, $\partial^2 \Phi/\partial \varphi^2>0$
for AFM state are satisfied for $\varphi=\varphi_0$,
$r'_{12}=r_{12}$, $r'_{13}=r''_{13}=r^{(0)}_{13}$,
$r'_{24}=r''_{24}=r^{(0)}_{24}$, as can be seen from the following
relations
\begin{eqnarray}\label{min_E_AF_13}
 && \frac{\partial \Phi}{\partial
  \varphi}={Na}\left[J'(r'_{12})\frac{2\sin(\varphi_0-\varphi)}{\sqrt{5-4\cos(\varphi-\varphi_0)}}+J'(r_{24})\frac{\sin\varphi(\cos\varphi_0-\cos\varphi)}{\sqrt{\sin^2\varphi_0+(\cos\varphi_0-\cos\varphi)^2}}
  \right.\\
  &&+\left.J'(r_{13})\frac{\cos\varphi(\sin\varphi-\sin\varphi_0)}{\sqrt{\cos^2\varphi_0+(\sin\varphi_0-\sin\varphi)^2}}\right]+2a^2K({\mathbf{k}_7})\sin(\varphi-\varphi_0)=0
  \nonumber
\end{eqnarray}

\begin{equation}\label{minAF_23}
  \frac{\partial^2 \Phi}{\partial
  \varphi^2}={Na}\left[J'(r_{24})\sin\varphi_0+J'(r_{13})\cos\varphi_0-2J'(r'_{12})\right]+2a^2K({\mathbf{k}_7})>0.
\end{equation}
Inequality (\ref{minAF_23}) is obviously satisfied due to the
already mentioned fact that the exchange integral is a positive
and monotonically decreasing function of intermolecular distance.

So, in AFM state the crystal lattice is stable with respect to
distortions even in the case of vanishingly small stiffness
$K({\mathbf{k}_7})$. As in the case of $\varepsilon$-phase, it is
\emph{the exchange forces} that keep the lattice from distortion.
In the phase with the long-range magnetic ordering the values of
the ``exchange bonds'' pulling the O$_2$ molecules in opposite
directions
 are equal (compare with $\varepsilon$-phase,
Fig.\ref{Fig_epsilon_def}), and this impedes nonsymmetrical
distortion of O$_8$-rhombuses.

It is also instructive to compare the magnetic energies of AFM
\begin{equation}\label{Energy_AF_2}
 E_{AFM}=N[J(2a\cos\varphi_0)+J(2a\sin\varphi_0)-2J(a)],
\end{equation}
and $\varepsilon$-phases
\begin{equation}\label{Energy_eps_2}
 E_{\varepsilon}=\frac{N}{2}[J(\sqrt{2}a)-3J(a)].
\end{equation}
It is obvious that  in a nondeformed hexagonal lattice
($\phi_0=60^\circ$) $E_{\varepsilon}< E_{AFM}$ for any value of
$a$. This means that the AFM state of $\alpha$ and $\delta$-phases
is stabilized by the long-range elastic forces that produce
homogeneous deformation (striction) of crystal lattice, as it was
shown in Ref.\onlinecite{Loktev:1981E}.

\section{Conclusions}
In summary, we have calculated the wave functions of singlet state
implemented on the 4(O$_2$) cluster and found that the exchange
energy of the ground state (\ref{wave_function singlet_1}) is
lower than that in another singlet states implemented on dimers,
trimers and quaters. Interactions between next-to-nearest
neighbors (i.e., between O$_2$ molecules located in the opposite
corners of rhombuses) plays an important role in stabilization of
the magnetic and crystal structure in the ground state.

We have shown that the observed distortion of crystal lattice and
formation of 4(O$_2$) quadrates in $\varepsilon$-phase can be
explained by strong magnetoelastic contribution into exchange
energy along with the softening of $u(\mathbf{k}_7)$ optical
phonon mode. Stability of the distorted lattice in
$\varepsilon$-phase is then due to antiferromagnetic character of
exchange interaction in solid oxygen.

The same magnetoelastic forces ensure stability of the AFM
long-range phases ($\alpha$, $\delta$) with respect to
inhomogeneous distortion of crystal lattice even in the case when
stiffness constant of $u(\mathbf{k}_7)$-mode is vanishingly small.

\end{document}